\documentstyle[prd,aps,twocolumn,epsfig]{revtex}
\addtocounter{equation}{9}
\begin{document}
\draft

\title{
\centerline{\bf Geometric Phase and Quantum Potential} }
\author{R. Dandoloff\cite{symbol2}}
\address{Laboratoire de Physique Th\'eorique et Mod\'elisation,
Universit\'e de Cergy-Pontoise, 5 Mail Gay-Lussac, 95031 Cergy-Pontoise
Cedex, France}

%
%
\abstract{We show that the geometric phase of L\'evy-Leblond
arises from a law of parallel transport for wave functions and
point out that this phase belongs to a new class of geometric
phases due to the presence of a quantum potential.}
 \maketitle2 \narrowtext

Since M. Berry \cite{1} has shown that the phase of the wave
function may play an important role, there have been many papers
investigating this effect, which now is widely known as Berry's
phase. J.-M. L\'evy-Leblond \cite{2} has discovered yet another
geometric phase. He has shown that a quanton that propagates
through a tube, within which it is confined by impenetrable walls,
acquires a phase when it comes out of the tube. For simplicity he
has considered a tube with square section of side $a$. The tube
has a length $L$. Before entering the tube the quanton's wave
function is: $\phi = e^{\frac{i}{\hbar}pz}$, where $z$ is the axis
of the tube and $p$ is its initial momentum. In the tube the wave
function has the following form:

$$\psi=sin(n_x\pi\frac{x}{a})sin(n_y\pi\frac{y}{a})e^{\frac{i}{\hbar}p'z}, \eqno(1)$$
with appropriate transverse boundary conditions. After entering
the tube the energy $E$ of the quanton is unchanged but satisfies:

$$E = \frac{p'^2}{2m}+({n_x}^2+{n_y}^2)\frac{\pi^2}{2ma}. \eqno(2)$$

For the simplest case: $n_x=n_y=1$, L\'evy-Leblond obtained the
following additional phase for the wave function, after the
quantum has left the tube:

$$\Delta\Phi=\frac{\pi^2\hbar^2}{pa^2}L. \eqno(3)$$

Subsequently R.E. Kastner \cite{3} has related this to the quantum
potential that arises in the tube. Now let us write  the wave
function in the tube in polar form:
$\psi_1=Re^{\frac{i}{\hbar}S+\frac{i}{\hbar}pz}$ (the eventual
changes in the phase of the wave function, due to the presence of
the tube, are now concentrated in $S$). In order to single out the
influence of the tube on the wave function, we will write the wave
function in the following form: $\psi_1=\psi
e^{\frac{i}{\hbar}pz}$. The quantum potential which corresponds to
$\psi$  is given by the following expression:

$$Q=-\frac{\hbar^2}{2m}\frac{\Delta
R}{R}=\frac{\pi^2\hbar^2}{ma^2}. \eqno(4)$$

Now, let us turn our attention to the laws of parallel transport
for the wave function, related to the appearance of the different
geometric phases. For the Berry's phase the law of parallel
transport for the wave function has been established by B. Simon
\cite{4}:

$$Im<\psi|\dot{\psi}>=0, \eqno(5)$$
where ($\,\,\dot{}\,\,$) stands for a time derivative. For a
spin-$\frac{1}{2}$ in magnetic field this law of parallel
transport is the analog of the Fermi-Walker parallel transport
\cite{5}.

For the L\'evy-Leblond's phase the law of parallel transport for
the wave function $\psi$ is given by the following expression:

$$ Im<\psi|\dot{\psi}>=-\frac{1}{\hbar}Q|\psi|^2. \eqno(6)$$

In L\'evy-Leblond's {\it gedanken experiment} the wave function
acquires an additional phase after the quanton has left the tube:

$$\psi(t+\Delta t)=e^{-\frac{i}{\hbar}Q\Delta t}\psi(t), \eqno(7)$$
which after expansion in $\Delta t$ leads to the law of parallel
transport: $Im<\psi|\dot{\psi}>=-\frac{1}{\hbar}Q|\psi|^2$
eqn.(6). Indeed:

$$ Q\Delta t = \frac{\pi^2\hbar^2}{ma^2}\Delta
t=\frac{\pi^2\hbar^2}{ma^2}\frac{mL}{p}=\frac{\pi^2\hbar^2}{pa^2}L
=\Delta\Phi \eqno(8).$$

If we use the polar form for the wave function, eq.(6) gives:
$\frac{\partial S}{\partial t}=-Q$. This means that this new law
of parallel transport eliminates the quantum potential from the
"quantum" Hamilton-Jacobi equation and we are left with the
classical Hamilton-Jacobi equation for a free particle. The whole
"quantum information" is now carried by the phase of the wave
function.

Now, after we have established the law of parallel transport for
$\psi$ in the case of the geometric phase of L\'evy-Leblond, we
see that the nature of this phase is quite different from the
Berry's phase. The appearance of the geometric phase of
L\'evy-Leblond is related to the presence or not of constraints in
the system. In this case the quanton is constrained to propagate
in a tube.

Let us now turn our attention to quite a different type of
constrained system and see how a new geometric phase, based on the
new law of parallel transport may appear there. We will consider a
quantum particle constrained to be on a circle \cite{6,7,8}. In
this case the wave function has the form: $\psi\sim
sin(\frac{ns}{\rho_0})$ ,where $\rho_0$ is the radius of the
circle and $s$ is the arc-length with origin at the tangent point.
In this way the wave function on the circle has a node at the
tangent point. In the case of a particle constrained to move on a
circle the value $n=\frac{1}{2}$ is also allowed \cite{7,8}. The
corresponding quantum potential for $n=\frac{1}{2}$ now is:

$$Q=\frac{\hbar^2}{8m\rho_0^2}. \eqno(9)$$

The quantum potential $Q$ is exactly equal to the constant $E_0$
which appears in the Hamiltonian for a particle on a circle with
radius $\rho_0$ following the Dirac quantization procedure for
constrained systems \cite{6}. The phase which would acquire a
quanton travelling along the circle is then:

$$Q\frac{2\pi\rho_0m}{p}=\frac{\pi\hbar^2}{4\rho_0p}. \eqno(10)$$

Note that if the circle becomes very small, then the geometric
phase can not get bigger than $\sim\frac{\hbar}{m}$. This limit is
imposed by the Heisenberg's uncertainty relation
$\rho_0p\sim\hbar$. This is not the case for the L\'evy-Leblond's
phase which can get very big provided $L\gg a$. One can imagine
the following {\it gedanken experiment} in order to test the
appearance of this new phase: a flow of quantons propagating on a
straight line with a circle of radius $\rho_0$ tangent to it. The
scattering at the point where the circle touches the straight line
will split this flow into two: one following the straight line and
the other following the circle and then the straight line (we are
not interested in the reflected flow which goes back). At the exit
these two flows will produce an interference picture which may
then be measured.

\begin{figure}
\begin{center}
\includegraphics[width=6cm]{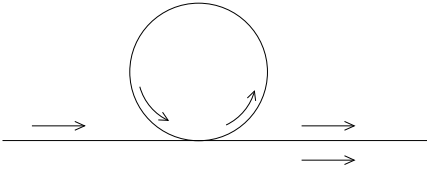}
\caption{A flow of quantons propagating along a straight line and
a tangent circle. } \label{newphase}
\end{center}
\end{figure}

As a conclusion, we have established the law of parallel transport
for the wave function in the case of L\'evy-Leblond's geometric
phase and have shown that this phase is not another manifestation
of Berry's phase. This phase is directly related to the constraint
imposed on the system. We have also shown that another constrained
system, a quantum particle on a circle, may exhibit the same kind
of geometric phase. The quantum potential gives a direct insight
into the problem why there is an additional constant appearing in
the Hamiltonian of this system. Acknowledgments:The author thanks
T.T.Truong for discussions.



\end{document}